\documentclass[smallcondensed,review,natbib]{svjour3}          % twocolumn

\usepackage{graphicx}

\usepackage[group-digits=true,group-minimum-digits=4,per-mode=symbol]{siunitx}    % to format numbers and units

\usepackage[driverfallback=dvipdfm]{hyperref}       % enhance documents that are to be output as HTML and PDF
\hypersetup{bookmarksopen=true,
                        bookmarksnumbered=true,   % number bookmarks
                        pdftitle={Comparison of Four Space Propulsion Methods for Reducing Transfer Times of Crewed Mars Mission},
                        pdfauthor={Andre Guerra, Orfeu Bertolami, Paulo Gil},
                        pdfsubject={Article comparing four different propulsion methods for a crewed mission to Mars},
                        pdfstartview=FitH,
                        pdfdisplaydoctitle=true,
                        pdfpagemode=UseNone} %UseOutlines

\usepackage{amsmath,amssymb}

\usepackage[para]{threeparttable}

\begin{document}

  %------------------------------------------------------------------------
  %	TITLE SECTION
  %------------------------------------------------------------------------
  \title{Comparison of Four Space Propulsion Methods for Reducing Transfer Times of Crewed Mars Mission}

  \author{Andr\'{e} G. C. Guerra\thanks{Corresponding author}\and
    Orfeu Bertolami\and
    Paulo J. S. Gil
  }

  \institute{Andr\'{e} G. C. Guerra \at
    Instituto Superior T\'{e}cnico, Universidade de Lisboa, Av.\ Rovisco Pais, 1049-001 Lisboa, Portugal \\
    \email{andre.gc.guerra@gmail.com} \\
    \emph{Present address:} CEiiA, Av. Dom Afonso Henriques, 1825. 4450-017 Matosinhos, Portugal.  %  if needed
    \and
    Orfeu Bertolami \at
    DFA, Faculdade de Ci\^{e}ncias, Universidade do Porto, Rua do Campo Alegre 687, 4169-007 Porto, Portugal \\
    \email{orfeu.bertolami@fc.up.pt}
    \and
    Paulo J. S. Gil \at
    CCTAE, IDMEC, Instituto Superior T\'{e}cnico, Universidade de Lisboa, Av.\ Rovisco Pais, 1049-001 Lisboa, Portugal \\
    \email{paulo.gil@tecnico.ulisboa.pt}
  }

  \date{Received: 07/04/2021 / Accepted: date}

  \maketitle

  \begin{abstract}
    We assess the possibility of reducing the travel time of a crewed mission to Mars by examining four different propulsion methods and keeping the mass at departure under \SI{2500}{\tonne}, for a fixed architecture. We evaluated representative systems of three different state of the art technologies (chemical, nuclear thermal and electric) and one advance technology, the ``Pure Electro-Magnetic Thrust'' (PEMT) concept (proposed by Rubbia). A mission architecture mostly based on the Design Reference Architecture 5.0 is assumed in order to estimate the mass budget, that influences the performance of the propulsion system. Pareto curves of the duration of the mission and time of flight versus mass of mission are drawn. We conclude that the ion engine technology, combined with the classical chemical engine, yields the shortest mission times for this architecture with the lowest mass and that chemical propulsion alone is the best to minimise travel time. The results obtained using the PEMT suggest that it could be a more suitable solution for farther destinations than Mars.
  \end{abstract}

  %-----------------------------------------------------------------------
  %	ARTICLE CONTENTS
  %-----------------------------------------------------------------------

  % ----------------------------------------------------------------------
  %  Introduction
  % ----------------------------------------------------------------------
  \section{Introduction}\label{sec:Introduction}

  Interplanetary space travel takes a long time with current technology. For example, it takes more than eight months to reach Mars in a minimum energy transfer and almost three years for a return mission~\citep{Wiesel2010sc-dyn}. In the case of crewed missions, such long travel times, and the long stay in Mars waiting for a low energy opportunity to return, require huge life support systems (food, air, etc.) capable of enduring the harshness of the space environment for long periods. Health problems due to long periods in microgravity (during travel) or low gravity conditions (in Mars) are well known, as well as the exposure to radiation and the risk of solar storms during travel~\citep{Cucinotta2009}, that increase with time and put at risk the success of the mission. These difficulties could be minimised with a considerable diminishing of both the travel time and the waiting time to return. Unfortunately, for moderate  gains in travel times, there is a penalty in the stay time imposed by the synodic period between the planets and the need to wait for their correct alignment.

  There have been several proposals to both cut the total mission or trips duration~\citep{Cassenti1998,Wooster2007}, or to limit the amount of propellant, usually with a penalty on time~\citep{Patel1998,Landau2006,Hughes2014}. Due to the exponential nature of the rocket equation~\citep{Wiesel2010sc-dyn}, going faster with the help of engines means more delta-v (i.e. propellant) and it can be advantageous not to go as fast as we would want to limit the mass, which is a proxy for money. One developed concept was the cycler orbits, that use fly-bys in the planets to save delta-v~\citep{Landau2004,Landau2006b}. These cycler orbit can be interesting for multiple missions~\citep{Pelle2019}, since the main transfer spacecraft (S/C) does not need to accelerate, only the one with the crew and support cargo. However, times-of-flight are high and the concept is more interesting as shuttle in the context of large and continuous service between the planets. An important way to limit delta-v is using aerocapture~\citep{MarsArchitectureSteeringGroup2009,Wright2006}. Using a fly-by in Venus to surpass the unfavourable planet alignment and cut the waiting time was also studied~\citep{Hughes2014}, but the main advantage is a smaller delta-v. Other more far side options have also been proposed, such as using a tether to assist the escape from the sphere of influence of the planets making the trip faster to up to 130 days or saving delta-v~\citep{Nordley2001}, but are not able to significantly reduce the mission duration, and the associated risk.

  Because of the difficulties of cutting the time short, the most developed mission proposals use solutions near the minimum energy. The probably most developed proposal, dubbed ``Design Reference Mission'' (DRM), was developed in 1992-1993 and was based on a concept by Robert Zubrin, the Mars Direct~\citep{Portree2001}. It has been revised several times, with the last version designated the Mars Design Reference Architecture (DRA) 5.0~\citep{MarsArchitectureSteeringGroup2009}. These concepts typically divide the mission in a cargo S/C, with a minimum energy transfer, sent in advance and a crewed S/C that can be a little faster. DRA 5.0 comprises a crew of six astronauts and transfers to and from Mars of about \SI{180}{\day} (for the crew). Together with a waiting time of about \SI{500}{\day}, the total mission would take around \SI{900}{\day}~\citep{MarsArchitectureSteeringGroup2009}. It requires a descent/ascent vehicle (DAV), a surface habitat (SHAB) and a multi-module Mars transfer vehicle (MTV), with each module sent to Mars weighing between \SI{60}{\tonne} and \SI{80}{\tonne}. It also uses nuclear thermal rockets as propulsion, as well as aerocapture (for the cargo S/C only) and in situ resource utilisation (ISRU). Other concepts were developed, typically variants of DRM with updated technology but without considerable differences in the mission duration, such as the ``Austere Human Missions to Mars''~\citep{Price2009}. Most concepts include insufficiently developed technologies, such as ISRU or aerocapture, increasing the risks, cost, or the development time of the mission.

  If a substantial increase in delta-v can be considered, however, it is in principle possible to solve both the slow trip and the long waiting time on Mars to return issues, imposed by the synodic period between the planets~\citep{Wertz2004b}. A sketch of a rapid mission with impulsive thrust has already been proposed by~\citet{Amade2010}, but with a total mission duration of about one year. All proposals requiring very high delta-v depend on the propulsion performance~\citep{Erichsen1997,Berend2013}. Typically, modern, or future, propulsion methods are required~\citep{Diaz2013} and some performance measure must be used to evaluate the merits of different alternatives~\citep{Erichsen1997}.

  In this paper, we compare the performance of different types of propulsion systems with the objective of evaluating the possibility of cutting short the total travel time of an interplanetary crewed S/C. We focus the analysis on a crewed mission to Mars as it is the next natural step for human exploration of the solar system. The problem is highly complex if all possible mission architectures are considered. Features such as ISRU or aerobraking can be important for a feasible solution, but they require a more detailed analysis about their implications and advantages on a rapid mission, as they have many implications in the architecture and the risk of the mission. Therefore, a mission architecture was selected in order to minimize mass (which means cost) without using such options. The goal is to establish a simple baseline, while different propulsion systems, in type and size, are tested to assess the impact on the duration of the mission. Further, more complicated, options are deferred for future work.

  Only the crewed component of the system has a strong requirement to minimise travel time. Therefore, we adopt a cargo plus crewed, two S/C, approach similar to DRA~\citep{MarsArchitectureSteeringGroup2009} and other missions, although without ISRU. To further minimise cost and mass, while increasing system's performance, modules of the S/C no longer needed are discarded as soon as possible, following the successful lunar orbit rendezvous approach used in the Apollo project~\citep{Woods2011}.

  Regarding our main variable, the propulsion system, we select four technologies for evaluation and determine the minimum transfer time possible for the crewed component, as a function of the size of the engines, i.e. thrust (respecting known engine size constraints). Selected propulsion technologies include: classical chemical engines (as it has been the workhorse of space exploration), nuclear thermal engines, modern electrical engines and a more conjectural concept of the many that have been proposed as the next revolutionary engine. We selected as the revolutionary engine the ``Pure Electro-Magnetic Thrust'' (PEMT) concept because it promised, in theory, to be highly effective due to its full conversion of mass into energy with momentum usable for thrust. We are interested in finding the minimum order of magnitude of the mission duration, within feasible bounds, and compare the merits of the different propulsion systems. We have not optimised the mission for each propulsion system. We have instead considered only transfer trajectories that provide representative performance for each system.

  The remainder of this paper is structured as follows: after review of the selected propulsion systems to establish the propellant consumption and how they can be scaled for more thrust, we define the features of baseline mission architecture and establish the type of transfer trajectories to use. We analyse the required mass determined by the propulsion against the mission time and discuss the results.

  % ----------------------------------------------------------------------
  %  Modern Propulsion Systems
  % ----------------------------------------------------------------------
  \section{Selected Propulsion Technologies}\label{sec:Propulsion_Systems}

  Many concepts for crewed missions to Mars resort to chemical propulsion or nuclear thermal systems but do not include the most advanced types of propulsion systems, including new chemical engines~\citep{Kemble2006,Turner2008,Vulpetti2008}. We have selected four examples of state of the art or advanced propulsion technologies and evaluate their potential to reduce the duration of human Mars missions (similar to the DRA 5.0 architecture), respecting a reasonable criteria for the total mass of the mission of less than \SI{2500}{\tonne}.

  The selected propulsion systems exhibit high specific impulse ($I_{sp}$), high thrust ($F_T$) and high thrust-to-weight ratio ($F_T/w$) among the numerous technologies available in the literature. We have considered propulsion technologies currently being tested or with flight proven capabilities. In addition, one propulsion system was examined that is an exception to this latter premise, the PEMT, presented by Carlo Rubbia~\citep{Rubbia2002}. For a broader discussion of propulsion systems, including a putative gravity control, the reader is referred to references~\citep{Turner2008,Bertolami2002,Tajmar2004,Bertolami2006,Sankaran2004}.

  The four propulsion systems considered are the:
  \begin{enumerate}
    \item Common Extensible Cryogenic Engine (CECE) -- representing the classical chemical propulsion;
    \item Nuclear Engine for Rocket Vehicle Application (NERVA) II -- assumed to be representative of nuclear thermal propulsion technology;
    \item Radio Frequency Ion Technology (RIT) XT -- assumed to be representative of modern electric propulsion technology;
    \item Pure Electro-Magnetic Thrust (PEMT) -- an advanced concept propulsion concept.
  \end{enumerate}

  % ----------------------------------------------------------------------
  %  Classical Chemical Engines
  % ----------------------------------------------------------------------
  \subsection{Classical Chemical Engines}\label{subsec:Chemical_Engines}

  In this brief discussion we do not consider storable, monopropellant, or solid systems~\citep{Kemble2006} but only cryogenic systems, due to their higher thrust (when compared to the aforementioned systems) and $I_{sp} \approx \SI{400}{\second}$~\citep{Turner2008}.

  Chemical engines present high thrust and relatively low $I_{sp}$, and accelerating continuously with a chemical engine would rapidly lead to an impractical mass budget. They are, therefore, used as impulsive engines, i.e.\ only active for small intervals of time (and the S/C is in free fall for most of its trajectory). Nevertheless, chemical engines can be crucial in escape and capture manoeuvres, because of their high thrust. Furthermore, impulsive engines can minimise the so called gravity losses for the same total delta-v available, by being used at once in a more favourable location~\citep{Oberth1972} --- usually deeper in the gravity well --- and with the additional advantage of possibly dumping empty propellant tanks sooner, maximising performance.

  The most important limitation of the chemical propulsion is that it is limited to the available chemical energy and thermodynamic conditions of the propellants~\citep{Turner2008}. The main point of developing other propulsion methods is to overcome these limitations.

  For our Mars mission, we only consider operations in space, taking for granted that some launch vehicle takes the system into orbit. An engine with the ability to be restarted is also required, to cope with the various mission phases. Examples of modern chemical engines fulfilling these criteria are the Vinci engine, designed for the upper stage of Ariane~5~\citep{EADSAstrium2012}, the RL10B-2, the latest version of the RL~10 engine used in the Delta~IV launch vehicle~\citep{Pratt_WhitneyRocketdyne2009_RL10B} and the Common Extensible Cryogenic Engine (CECE) of Pratt \& Whitney Rocketdyne~\citep{Pratt_WhitneyRocketdyne2009_CECE}, also an evolution of the RL~10. Of course, forthcoming technological developments involving, for instance, zero boil off for cryogenic propulsion fluids might be considered but as most often these represent improvements rather than breakthroughs they will probably not affect considerably our assumptions and conclusions.

  We selected as representative of the chemical propulsion system the CECE engine. Although the thrust of the Vinci engine (\SI{180}{\kilo\newton}) is higher than the others ($\approx \SI{110}{\kilo\newton}$), it has a much higher mass (almost the double of the lighter). Additionally, they all have similar $I_{sp}$. Consequently, the CECE has a much higher thrust-to-weight ratio, which, together with the ability to restart many more times (50, instead of 5 or 15), makes it the best choice for such a complex mission. Relevant characteristics of one CECE engine can be found in Table~\ref{tab:Engines_Global}, where in all cases the mass of propellants and their deposits is not included since they can be discarded and are accounted separately.

  \begin{table}[!hbt]
    \centering
    \begin{threeparttable}
      \caption[Engines Selected]{Main characteristics of selected engines for a system consisting of one engine (data from~\citet{Pratt_WhitneyRocketdyne2009_CECE,Black1964,Leiter2001}).
        The mass of propellants and their deposits is not included}
      \label{tab:Engines_Global}
      \begin{tabular}{cccccc}
        \hline\noalign{\smallskip}
        Engine      & Power [\si{\watt}]  & $I_{sp}$ [\si{\second}]   & Thrust [\si{\newton}] & Mass [\si{\kilogram}]   & $F_T/w$ [\si{\newton\per\kilogram}] \\
        \noalign{\smallskip}\hline\noalign{\smallskip}
        CECE        & -                   & 465                       & \num{0.11e6}          & 256                     & 435 \\ %\citep{Pratt_WhitneyRocketdyne2009_CECE}
        NERVA II    & \num{5.0e9}         & 785                       & \num{1.0e6}           & \num{34e3}\,\tnote{a}   & 30 \\ %\citep{Black1964}
        RIT-XT      & \num{3260}          & \num{4600}                & 0.12                  & 32\,\tnote{b}           & \num{3.7e-3} \\ %\citep{Leiter2001}
        PEMT        & \num{6.1e9}         & \num{30.6e3}              & 20                    & \num{32e3}\,\tnote{c}   & \num{0.64e-3} \\
        \noalign{\smallskip}\hline\noalign{\smallskip}
      \end{tabular}
      \begin{tablenotes}
        \item [a] The reactor's mass is \SI{11e3}{\kilogram}.\\
        \item [b] Includes mass of the required solar panels and power processing electronics.\\
        \item [c] Includes mass of the radiator and reflector.
      \end{tablenotes}
    \end{threeparttable}
  \end{table}

  % ----------------------------------------------------------------------
  %  Nuclear Thermal Engines
  % ----------------------------------------------------------------------
  \subsection{Nuclear Thermal Engines}\label{subsec:Nuclear_Engines}

  Nuclear thermal engines have been developed since the 1940s and were even considered for the upper stage on the Nova rocket (for the lunar direct launch mission)~\citep{Turner2008}. Their working principle is similar to chemical engines, with high thrust and $I_{sp} \approx \SI{800}{\second}$ and, therefore, should be treated as impulsive. A single propellant, usually hydrogen, is heated by the nuclear core and expelled through a nozzle while expanding. The core, usually an uranium derivative (like dioxide or carbide) or plutonium, releases heat due to the nuclear reaction, providing energy to the gas expansion and resulting in an $I_{sp}$ approximately the double of the chemical engines. The heat released is limited by the melting point of the materials~\citep{Kemble2006}.

  One of the engines with highest power to be developed within the Nuclear Engine for Rocket Vehicle Application (NERVA) program was the NERVA~II, which had the goal of achieving a higher $I_{sp}$ and thrust, with a lower weight than previous models~\citep{Black1964}. It was intended to serve as the propulsion system for crewed interplanetary missions with masses close to \SI{1000}{\tonne}. NERVA~II produced the required power with a uranium inventory of \SI{360}{\kilogram}, and had \SI{2}{\metre} in diameter~\citep{Black1964}. Temperatures of the hydrogen propellant could reach \SI{2755}{\kelvin}~\citep{Black1964}. One of the requirements of the program was an endurance of over \SI{600}{\minute}~\citep{Klein1970}. Its main features are shown in Table~\ref{tab:Engines_Global}.

  Another engine, with interesting features, developed within Project Rover program, was the Phoebus II engine, which featured $F_T/w = 38$ and $I_{sp} = \SI{790}{\second}$~\citep{Sue1987}. These values are a little higher but the difference to NERVA II is not too large and this engine showed overheating problems during the tests conducted at the time.

  Options such as the ORION, a nuclear pulse propulsion system developed in the 1950/1960s, was not considered due to its low $F_T/w$ (between one and six), and the need to blast nuclear material~\citep{Nance1965}. Notice that even though the ORION proposal has a greater $F_T/w$ than electric propulsion, the $I_{sp}$ of the latter is much larger than the former.

  % ----------------------------------------------------------------------
  %  Modern Electric Engines
  % ----------------------------------------------------------------------
  \subsection{Modern Electric Engines}\label{subsec:M_Electric_Engines}

  Electric propulsion overcomes the limitations of chemical engines by separating the energy source from the propellant material, and by not using thermodynamic mechanisms to accelerate particles. Common sources of energy are solar, nuclear power and radioisotope thermal generators (RTG)~\citep{Kemble2006}. The thrust produced by current technology is very small, when compared to chemical engines. However, they can achieve much larger $I_{sp}$, allowing the engine to run for longer periods with less propellant. The electric engine is treated as a non-impulsive engine since it is working most of the trajectory. In these systems we have to take into account not only the mass of the engine but also the mass of the associated energy system, e.g.\ solar panels and the power supply and control unit (PSCU)~\citep{Brophy2011}.

  The selected power source is the solar photovoltaic, since nuclear power systems represent a large increase of the engine mass, and the RTG technology can only achieve specific powers of \SI{5}{\watt\per\kilogram} (and are under development)~\citep{Kemble2006}. The size and mass of the required solar array can be estimated for the power level of the engine used (as indicated in Table~\ref{tab:Engines_Global}), and must be taken into account on the mass budget of the mission. We considered an efficiency of \SI{29.5}{\percent} for the solar cells, a density of $\rho_A^{sp}=\SI{0.84}{\kilogram\per\metre\squared}$, with degradation rate of \SI{0.4}{\percent} per year, and the worst case scenario of a three-year mission~\citep{EmcoreCorporation2011}. For the PSCU a direct-drive concept was used, with a corresponding mass, in \si{\kilogram}, given by a scaling parameter ($M^{PSCU} = 0.35 W + 1.9$, where $W$ corresponds to the input power in \si{\kilo\watt})~\citep{Brophy2011}. These values give a density for the associated energy system of about \SI{7}{\kilogram\per\kilo\watt}, well within what was used in other studies~\citep{Sankaran2004}.

  Several electric propulsion technologies are in use such as the arcjet, Hall effect, and gridded ion thrusters~\citep{Kemble2006}. Modern examples of electric engines include: NASA's Evolutionary Xenon Thruster (NEXT), an evolution of the already tested NASA Solar Technology Application Readiness (NSTAR) used in the Deep Space 1 mission~\citep{Emhoff2006} and Dawn~\citep{Brophy2003a}; the radio frequency ion thruster RIT-XT, which works by generating ions using high frequency electromagnetic fields and is very similar to the RIT-22 engine (in design and thrust-to-weight ratio) but with higher $I_{sp}$~\citep{Leiter2001}; and the PPS~1350-G, which is a plasma thruster with flight proven capability (SMART-1 mission)~\citep{Snecma2011}.

  To select the electric propulsion system, including all elements required such as power processing electronics and power source, we considered not only the specific impulse $I_{sp}$ but also the thrust-to-weight ratio, because the latter also affects the transfer time~\citep{Miele2004}, which is our main concern. Moreover, the power requirements of the engine also influence the engine selection, since it has an impact on the size and mass of the associated energy system. Consequently, the selection of the electric engine proved to be more difficult than the other cases.

  We evaluated both the PPS~1350-G and the RIT-XT engines in simulated transfers to Mars. Differences in performance were found to be small, with an apparent advantage to the latter, and we end up deciding for the RIT-XT engine as the representative engine of the electric propulsion. Its main characteristics are shown in Table~\ref{tab:Engines_Global} (for a single RIT-XT engine, including the energy system).

  Using the RIT-XT engine as baseline, we also briefly discuss the possible gains in performance if some future technologies would enhance the specific impulse or the thrust-to-weight ratio (see section~\ref{subsec:RD_Electric_Engines}). We considered a specific impulse and a thrust-to-weight ratio more than two times and three times the corresponding value of the RIT-XT engine, respectively. While the considered specific impulse increase would correspond to a direct technology enhancement in the engine, the increase in the thrust-to-weight ratio was based in foreseen developments in the energy system.

  Another electrical propulsion technology that has been discussed for crewed Mars missions is the magnetoplasmadynamic thruster (MPD)~\citep{Sankaran2004,Ageyev1993}. The expected total thrust generated by these engines is orders of magnitude larger than the previously discussed electrical engines, even though the thrust-to-weight ratio and specific impulse is of the same order. However, the MPD that is claimed to have better results~\citep{Sankaran2004}, explains that the engine could work only for about 1000 hours before degradation occurs~\citep{Ageyev1993}, which is considerably less than the expected trip times. Furthermore, the $F_T/w$ and $I_{sp}$ values for the MPD are well within what we test for possible future technologies in section~\ref{subsec:RD_Electric_Engines}. Consequently, we decided not to explicitly consider MPD thrusters as continuous impulse engines in our study, although with the expected levels of thrust of these engines they could be considered to be in the frontier between impulsive and non-impulsive engines, and future studies should be performed (not only to increase the endurance time but also to explore what can be achieved with these options).

  % ----------------------------------------------------------------------
  %  Nuclear Pure Electromagnetic Thrust
  % ----------------------------------------------------------------------
  \subsection{Nuclear Pure Electromagnetic Thrust}\label{subsec:M_PEMT_Engine}

  Propulsion technologies based on new concepts are constantly being proposed and tested, as is the case of Electrodynamic Tethers, MagSails, Plasma Sails, and Solar Sails~\citep{Vulpetti2008}. We included the Nuclear Pure Electromagnetic Thrust (PEMT) concept in our evaluation to see if such an advanced concept had the potential for significantly reduce flight times for human missions to Mars. This engine uses the momentum of emitted photons, instead of expelling a working fluid to create thrust~\citep{Rubbia2002}.

\begin{figure}[!hbt]
    \centering
    \includegraphics[width=1\textwidth]{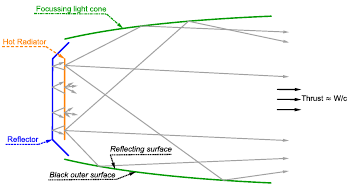}
    \caption[Rubbia's Concept Engine]{Scheme of Rubbia's engine concept~\citep{Rubbia2002}}
    \label{fig:Rubbia_Engine}
  \end{figure}

  The thermal energy produced in a nuclear reactor is used to heat a radiator, which emits electromagnetic radiation (photons). The radiator is in front of a reflecting surface to direct the radiation that produces thrust. A Winston cone (Fig.~\ref{fig:Rubbia_Engine}), a non-focusing reflecting conical structure, can be used to collimate the radiation, resulting in a total thrust of $F_T=W/c$, where $W$ is the power and $c$ the speed of light~\citep{Rubbia2002}.

  The power emitted is related to the area of the surface of the radiator ($S$) and its temperature ($T$) through Stefan-Boltzmann's law, and the power produced by the nuclear engine. It is possible to ensure that a reasonable sized radiator yields say, \SI{20}{\newton}, for a radiator temperature of about \SI{3300}{\kelvin} (close to the boiling point of a coolant). Among the materials that can withstand this radiator high temperature, without melting, carbon nanotubes are the lightest (as suggested by~\citet{Rubbia2002}). The cone reflector must then envelop the radiator, and should have high reflectivity for the wavelength the radiator is emitting. For our radiator temperature, the reflector cone should be capable of reflecting visible and infrared radiation to be effective. Current technology for solar sails use composite booms (on the support structure) and Aluminised Mylar sails (or carbon fibre sail substrate), with densities of $\SI{10}{\gram\per\metre^2}$ (including the support structure)~\citep{Vulpetti2008}. Combining the densities of the materials and sizes of the structures, the radiator and reflector mass can be extracted, using the ideas of~\citet{Rubbia2002}.

  As power source we have selected a NERVA-like reactor, resized and improved, as this is one of the discussed nuclear reactors with a power level closer to our intended figure. It is assumed a ``NERVA 2000'' reactor which can produce \SI{22}{\percent} more energy with an increase in mass of \SI{26}{\percent}, to get a net \SI{20}{\newton} engine. The need for a nuclear reactor is justified by the $\SI{50}{\kilo\metre^2}$ of solar panels, weighting \SI{44}{\kilo\tonne}, required to produce the same power using the already mentioned technology~\citep{Vulpetti2008,Brophy2011}.

  In the PEMT concept no mass is expelled. However, it is possible to determine its $I_{sp}$, and compare it with the propellant mass spent by chemical thrusters~\citep{Rubbia2002}. If we assume that the fraction of mass transformed into energy through nuclear fission is $\xi = 10^{-3}$~\citep{Rubbia2002}, and it is ejected at the speed of light, $c$, then the effective exhaust speed is $v_{ex}=\xi c$. Thus, the specific impulse is given by $I_{sp} = \SI{30600}{\second}$. The main features of the engine are shown in Table~\ref{tab:Engines_Global}.

  The PEMT engine is, nevertheless, still a theoretical concept and presents many potential difficulties to be implemented in practice: control of nuclear reactions in space (in free fall cooled only by radiation), insulation of the engine and the rest of the S/C, transfer of energy from the reactor to the radiator, etc. The required high temperature is another challenge since the radiator material must withstand it without melting. Some solutions such as discussed by~\citet{Rubbia2002}\ are somewhat difficult to implement. We nevertheless decided to include this technology, and use its optimistic characteristics at face value, to assess if such an engine proposal would offer some real advantages over other well known options.

  % ----------------------------------------------------------------------
  %  Propellant Calculation
  % ----------------------------------------------------------------------
  \subsection{Propellant Calculation}\label{subsec:Fuel_Calculations}

  For engines that can be treated as impulsive, we consider the usual approximation of instantaneous change in velocity $\Delta v$, obtaining the propellant mass ($M_P$) through the Tsiolkovsky rocket equation. The complete mission trajectory requires $N$ manoeuvres that will be determined from the last to the first, since the propellant mass depends on the initial mass for each segment. The finite burn losses are taken into account by defining a loss factor~\citep{Brown1998}.

  For continuous thrust systems the spent propellant is determined at each instant by $\dot{m} = - {F_T}/{I_{sp} g_0}$, and taken into account in the numerical integration of the equations of motion.

  In the case of PEMT, the propellant required is the nuclear material for the reactor, which is not expelled (it yields photons). Therefore, we need to compute the amount of nuclear material to load the reactor with.

  The thrust force of PEMT is given by the radiated power ($W_{rad}$) divided by the speed of light ($F_T = W_{rad}/c$)~\citep{Rubbia2002}. We assume that the radiated power is equal to the power generated by the reactor, i.e. $W_{rad} = W_{reactor}$. Combining this with the time of flight (ToF), which is equal to the operational time of the reactor, $t_{on}$, we compute the total burnup of the reactor, $B = W_{reactor} \times t_{on}$, expressed in $\si{\giga\watt} \times \si{\day}$~\citep{Rubbia2002,Lamarsh2001}.

  The specific burnup, $sb$, of a nuclear material is the total energy released per unit of mass of nuclear propellant, and is expressed in $\si{\mega\watt} \times \si{\day\per\tonne}$. This is proportional to the fractional burnup $\beta$, defined as the ratio of the number of fissions for a specified mass of propellant to the total number of heavy atoms~\citep{Lamarsh2001}. Furthermore, if all propellant atoms where fissioned, $\beta = 1$, this would lead to \SI{950}{\giga\watt\day\per\tonne} for the uranium isotope 235 ($^{235}U$)~\citep{Lamarsh2001}. Consequently, the specific burnup is $sb = 950\beta\si{\giga\watt\day\per\tonne}$. Throughout our work we have used $\beta$ equal to \SI{4}{\percent}~\citep{Lamarsh2001}.

  To determine the mass of uranium, $M_{U}$, required to produce the needed energy, we divide the total burnup obtained, by the specific burnup, i.e. we use $M_{U}\,[\si{\tonne}] = {B\,[\si{\giga\watt\day}]}/{sb\,[\si{\giga\watt\day\per\tonne}]} = {W_{reactor} \times t_{on}}/{950 \beta}$. If the nuclear reactor is loaded with uranium dioxide ($UO_2$), enriched to almost weapon's grade (\SI{80}{\percent} of $^{235}U$), using $M_{UO_2} = M_{U} \times A_w^{UO_2}/A_w^{U}$ (where $A_w^{X}$ is the atomic mass of element $X$), we can compute the needed mass of uranium dioxide $M_{UO_2}$~\citep{Lamarsh2001}.

  % ----------------------------------------------------------------------
  %  Mission to Mars
  % ----------------------------------------------------------------------
  \section{Mission to Mars}\label{sec:Mission_Mars}

  The mass and time required for a mission to Mars is dependent on its architecture. We consider a mission similar to the DRA 5.0 concept, that includes a crewed S/C with a crew of four astronauts and an uncrewed, or cargo, S/C.

  The crewed S/C is comprised of a human habitat module (which houses the crew during transit to and from Mars, and includes all life support systems for the mission), a propulsion system and a transport capsule. The first two might be assembled in orbit. The transport capsule carries the crew from low Earth orbit (LEO) to the main S/C on Earth, and between the main S/C and low orbit on Mars (upon arrival).

  The uncrewed cargo S/C consists of the propulsion system, the payload for Mars operations, including the descent and ascent vehicle, and the propellant required for the astronauts to return to the Earth, that will be transferred to the main S/C while in Mars orbit.

  % ----------------------------------------------------------------------
  %  Mission Architecture \& Mission Timeline
  % ----------------------------------------------------------------------
  \subsection{Mission Architecture and Mission Timeline}\label{subsec:Mission_Architecture_Timeline}

  The same architecture is used to compare the performance of the different propulsion systems. Considering that the objective is to minimise the travel time for the astronauts, the crew and most of the cargo are sent separately. This allows sending cargo through a slower and more economic way, reducing substantially the initial mass of the crewed S/C.

\begin{figure}[!hbt]
    \centering
    \includegraphics[width=1\textwidth]{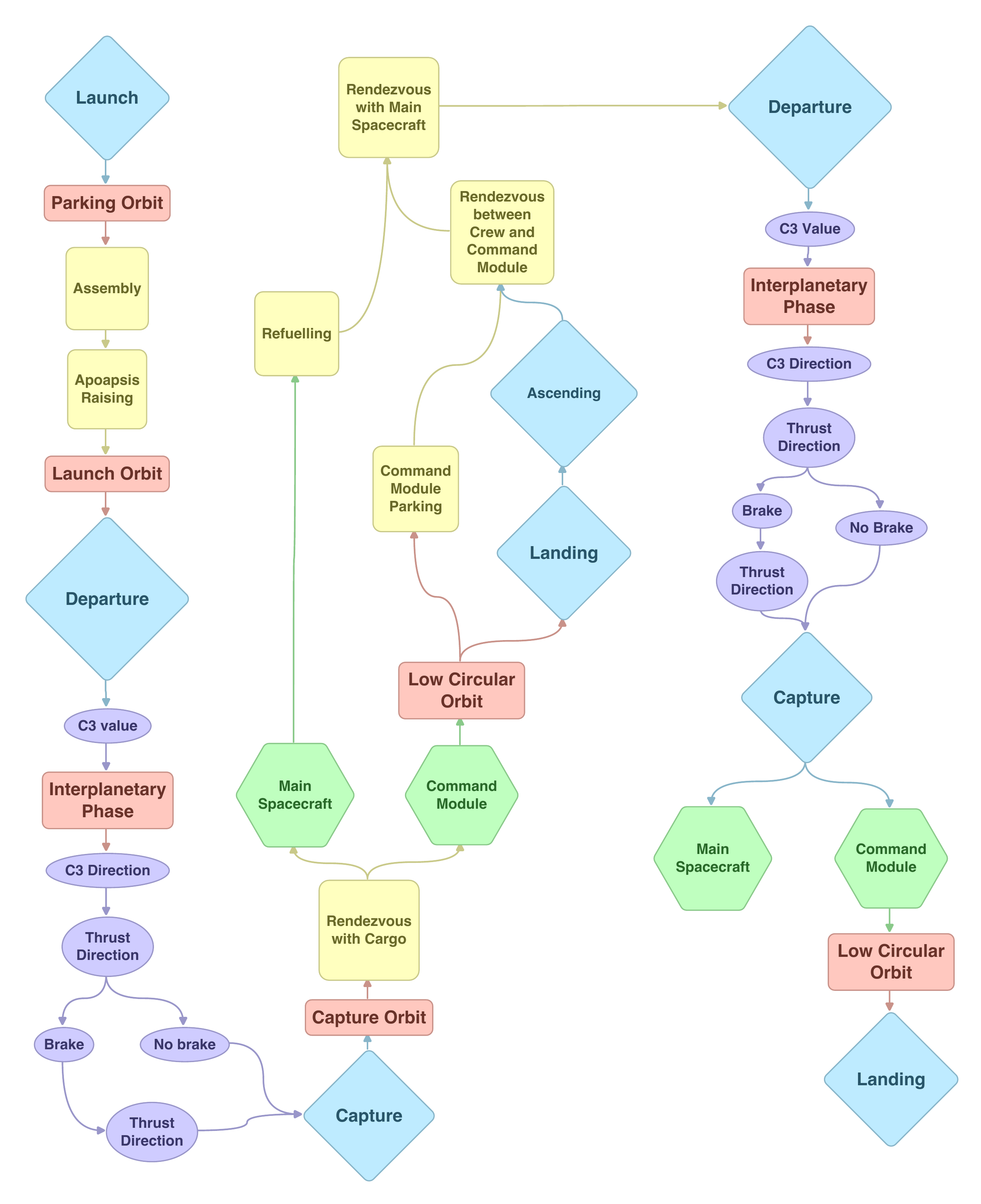}
    \caption[crewed Mission Timeline]{Sketch of the crewed mission timeline. Diamonds, represent \textit{Single instants of time}; rectangles, \textit{Mission phases}; squares, \textit{Actions}; oval shapes, \textit{Variable parameters} (C3 equals the square root of the departure velocity)}
    \label{fig:Mission_Timeline}
  \end{figure}

  The mission timeline for the crewed phase is displayed in Fig.~\ref{fig:Mission_Timeline}.

  % ----------------------------------------------------------------------
  %  Parking Orbit \& Departure
  % ----------------------------------------------------------------------
  \subsubsection{Parking Orbit and Departure}\label{subsec:Parking_Departure}

  We have considered that the mission starts at LEO, where a launch system can deliver all the required modules for the mission. Both the main (crewed) and the cargo S/C can be assembled at an initial circular parking orbit at about \SI{500}{\kilo\metre}, similar to the International Space Station, and which is high enough to allow assembly without decay.

  After assembling and testing, the S/C would raise its apogee to an orbit with eccentricity $0.9$, in a series of impulsive manoeuvres at perigee, so to minimise finite burn losses when compared to a direct interplanetary injection from LEO~\citep{Brown1998}. This saves mass by discarding the propellant tanks already used. The final elliptical orbit is selected to maximise its energy (but not too close to the escape energy to avoid complication with orbital perturbations and long periods of revolution). At the same time, we keep the perigee low to use the effectiveness of the Oberth effect when escaping from the Earth influence.

  If continuous thrust (when applicable) would be used at departure, too much time would be required to escape, since the continuous thrust considered is yet of relatively small intensity. Therefore, we only considered chemical (or nuclear thermal if applicable) propulsion for escape and capture manoeuvres, and for the apogee raising initial manoeuvre. In this context, we considered several escape velocities $\vec{v}_\infty$ leading to different initial velocities for the interplanetary phase.

  % ----------------------------------------------------------------------
  %  Interplanetary Transfer
  % ----------------------------------------------------------------------
  \subsubsection{Interplanetary Transfer}\label{subsec:Interplanetary}

  The proximity of Mars implies that interplanetary transfer solutions with fly-by manoeuvres will increase the travel times making these uninteresting for the main S/C. For simplicity, we did not consider this option for the cargo mission as well. We are mainly interested in comparing the relative performance of the different propulsion systems, whereas other, better options, for the cargo S/C can be used by all, improving the considered alternatives equally.

  For the impulsive-type propulsion, chemical and nuclear thermal, once the initial velocity is defined a coast trajectory (Lambert arc) takes the S/C to the arrival planet, where it is captured. For the continuous thrust propulsion, electric and PEMT, the direction and intensity of the thrust can make a considerable difference and must be determined in each case to obtain a suitable solution. The continuous thrust can also be allowed to brake because it could imply a smaller requirement of propellant for the capture.

  In any case, as our goal was to compare the relative performance of the propulsion systems, a full trajectory optimisation for each interplanetary orbit solution was not performed. Therefore, for the continuous thrust solutions a constant angle between thrust and velocity vectors was assumed, as explained in section~\ref{subsec:Trajectory}. For the impulsive engines, we also did not consider deep (impulsive) space manoeuvres as these are usually useful to synchronise with planets for fly-by, and they seem not to offer any advantage as compared with a larger delta-v near Earth (initial or final manoeuvre), due to the Oberth effect.

  % ----------------------------------------------------------------------
  %  Capture on Mars
  % ----------------------------------------------------------------------
  \subsubsection{Capture at Mars}\label{subsec:Capture}

  The capture manoeuvre at Mars is performed using the classical single impulse brake, since continuous thrust brake would require too much time to execute for its level of thrust~\citep{Rubbia2002,Kemble2006}. We also did not consider aerobraking for the crewed S/C for several reasons: it would be of high risk~\citep{JillL.H.PrinceRichardW.Powell}; it would require shielding and a somehow compact and strong architecture, adding to the mass and loosing part of its appealing (apart from a possible incompatibility with a high area of solar panels in some options); given the possible energies involved, it could be insufficient for the capture; and it would possibly have to be complemented with thrust. In this context of high arrival energies, aerobraking requires a whole different analysis beyond the scope of this work.

  The capture orbit, where the main S/C will be parked, has a periapsis altitude of \SI{300}{\kilo\metre}, selected by comparison with other missions~\citep{Fletcher2009,Graf2005}~and eccentricity $0.9$. We selected this high energy orbit such that its low periapsis allows for an effective capture manoeuvre under high velocity but minimising the propellant consumption during the capture. The main S/C is also used in the return trajectory and propellant is saved by not lowering the apoapsis. The propellant for the return trajectory is transported by the cargo S/C (sent previously and with the same capture orbit), and can now refill the crewed S/C for the return.

  We did not consider ISRU, except for recycling of air and water (possibly included into the capsule), as our architecture is adequate for a fast first mission, and ISRU presents a new set of challenges. Apart from the risk, use of ISRU in a large scale to grow food and obtain mission support resources in general only makes sense for prolonged stays, which is not applicable in the case we are considering. The only remaining interesting use of ISRU is for generating return propellants. However, in the present case, they would have to ascend into orbit (with a 0.9 eccentricity), and demanding a much larger means of transportation than just bringing astronauts back, given the high delta-v involved in the type of mission considered here. This will possibly erode the advantage of ISRU, as the achieved gain in propellant mass would have to compensate a considerable increase in the mass of the transport vehicle, and the increase in propellant mass required to transport this vehicle to Mars. Therefore, it is not certain that ISRU would bring considerable advantages for the considered type of mission and, although its analysis is beyond the scope of this work, it remains an open issue.

  The considered mission can be seen, thus, as a baseline simple mission, and further studies can be done to better investigate more sophisticated option, possible including ISRU or aerobraking. The above discussion suggests that the gains possibly offered by such technologies will not be of orders of magnitude for this type of missions, and qualitative conclusions will be unaltered.

  Meanwhile the crew (using the transport capsule) would get all the required payload from the the cargo S/C, and descend to a \SI{300}{\kilo\metre} circular orbit. Afterwards, the crew changes to the Mars operations vehicle, that will descend to the surface, while the transport capsule waits in orbit.

  % ----------------------------------------------------------------------
  %  Return Trajectory
  % ----------------------------------------------------------------------
  \subsubsection{Return Trajectory}\label{subsec:Return}

  After completion of the ground operations, the crew ascends from the surface in the Mars operations vehicle (or part of it), returning to the transport capsule at a low Mars orbit. After rendezvous and discard of the Mars operations vehicle, the transport capsule raises its orbit to return to the main S/C for the MtE (Mars to Earth) injection manoeuvre.

  The capture at Earth is similar to the one at Mars (the entire S/C is captured and possibly reused; also, propulsion systems with nuclear material would have to be carefully decommissioned). Subsequently, the crew enters the transport vehicle and returns to LEO where it will be transported back to Earth's surface.

\begin{figure}[!hbt]
    \centering
    \includegraphics[width=1\textwidth]{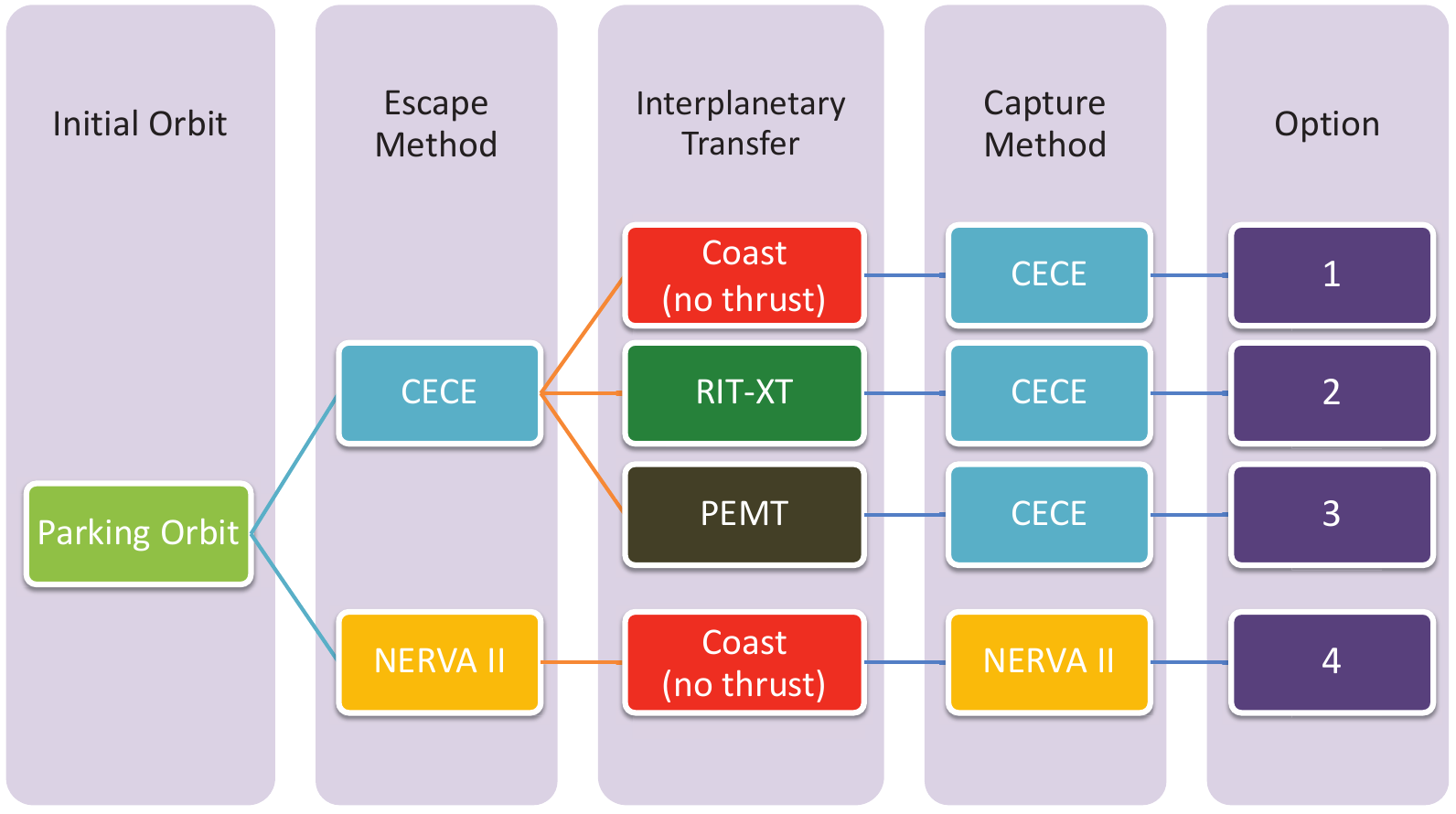}
    \caption[Mission Trade Tree]{Different engine configurations assumed for the crewed S/C}
    \label{fig:Architecture_Main}
  \end{figure}

  A mission trade tree is displayed in Fig.~\ref{fig:Architecture_Main}, where we can see which propulsion system is used in each mission segment. When a coast transfer is selected no propulsion system is used during the interplanetary phase.

  % ----------------------------------------------------------------------
  %  Cargo Mission Phase
  % ----------------------------------------------------------------------
  \subsubsection{Cargo Mission Phase}\label{subsec:Cargo_Phase}

  Starting from the same assembly circular LEO, the cargo S/C raises the apogee altitude, similarly and for the same reasons than the crewed S/C, to the same high elliptical orbit of eccentricity $0.9$, and performs an EtM (Earth to Mars) injection. Continuous propulsion could be used for this purpose advantageously. As the cargo S/C mass has mostly the same impact on the four discussed propulsion systems, we selected, for simplicity and without any adverse effects on the relative results, an impulsive (purely chemical) transfer approximated by a Hohmann transfer, especially since a fast trajectory is not required for the cargo (the main goal is to minimise the required energy saving mass).

  The cargo is to arrive at Mars before the crew, for the payload to be ready to descend to Mars. Upon arrival on Mars, the cargo S/C is captured to a highly elliptical orbit, similar to the crewed S/C, since the latter will have to be refilled with the propellants carried by the cargo S/C to return to Earth.

  % ----------------------------------------------------------------------
  %  Mass Budget
  % ----------------------------------------------------------------------
  \subsection{Mass Budget}\label{subsec:Mass_Budget}

  While crew, modules (habitat and cargo), and transport capsules remain constant, the type of propulsion and number of engines are variables of the problem. Propellants and tanks are also dependent on the execution time of the manoeuvres, since we considered different levels of finite burn losses~\citep{Brown1998}. The human habitat and cargo modules (including all necessary systems and payloads), and transport capsules, are an extrapolation from Mars Direct~\citep{Zubrin1996} and DRA 5.0~\citep{MarsArchitectureSteeringGroup2009} concepts using typical guidelines~\citep{Fortescue2011}.

  The mass budget for the mission is therefore comprised of:
  \begin{itemize}
    \item Crewed S/C (a fixed value including all systems and crew, except the propulsion system): $M_{fixed}^{Crewed} = \SI{38}{\tonne}$;
    \item Cargo S/C (again with exception of the propulsion system): $M_{fixed}^{Cargo}= \SI{42}{\tonne}$;
    \item Mars operations vehicle (which will follow with the cargo S/C) $M^{MOV}= \SI{18}{\tonne}$;
    \item Propulsion System, $M_{PS}$:
    \begin{itemize}
      \item Impulsive engines (chemical or nuclear);
      \item Continuous Thrust Engines, including:
      \begin{itemize}
        \item Electric propulsion case: solar panels including structure, and power processing electronics;
        \item Rubbia's concept engine case: the radiator \& reflector;
      \end{itemize}
    \end{itemize}
    \item Propellants, $M_{P}^T$;
    \item Propellant tanks, $M_{T}^T$;
  \end{itemize}
  The mass of the propellant tanks can be computed from the propellant mass (computed using the algorithm described in section~\ref{subsec:Fuel_Calculations}), using a fitting function from several specifications of existing tanks as input parameters~\citep{ASTRIUMSpaceTransportation2010}. Furthermore, they are sized in order to be empty and discarded after each large manoeuvre, as possible.

  Consequently, the dry mass of the crewed S/C is $M_{Dry}^{Crewed} = M_{fixed}^{Crewed} + M_{PS}$, where the tanks are not included since they are discarded after each manoeuvre. Its total mass is $M_{Total}^{Crewed} = M_{Dry} + M_{P}^T + M_{T}^T$. The dry mass of the cargo S/C is $M_{Dry}^{Cargo} = M_{fixed}^{Cargo} + M_{PS} + M_{R}^{FP}$. The propellant payload entry ($M_{R}^{FP}$) represents the propellants and tanks needed for the crewed return trajectory. The total cargo mass is $M_{Total}^{Cargo} = M_{Dry} + M_{P}^T + M_{T}^T$.

  Combining the total crewed S/C mass for the Earth to Mars transfer, with the total cargo mass (which includes the return propellant for the crewed S/C), yields the total mission mass ($M_{Total} = M_{Total}^{Crewed} + M_{Total}^{Cargo}$).

  % ----------------------------------------------------------------------
  %  Trajectory Problem \& Solution
  % ----------------------------------------------------------------------
  \subsection{Trajectory Problem and Solution}\label{subsec:Trajectory}

  To determine the trajectory and time of flight we adopted a simple patch-conic approximation, with a numerical integration in the interplanetary phase of the mission. We consider Earth and Mars to be in the same plane and in circular orbits around the Sun (with radius equal to the true semimajor axis). This makes the problem only dependent on the heliocentric angle between the planets and the characteristics of the mission, and not the specific launch date.

  The S/C escapes from, and is captured into, an elliptical orbit with instantaneous manoeuvres at the periapsis of the orbits, using chemical or nuclear thermal propulsion. The terminal velocity, $\vec{v}_\infty$, of the escape hyperbola can make any angle with the velocity of the departure planet, within the planet's orbital plane. The arrival velocity angle is determined by the interplanetary phase, a simple Lambert arc in the case of impulsive-type propulsion, or a trajectory determined by the continuous thrust that can include a brake segment to ease the capture manoeuvre.

  A relatively simple approach is used to obtain a solution of the continuous thrust trajectory. Continuous thrust has a high $I_{sp}$, and the expense of propellant is not the main issue, so to minimise the transfer time the propulsion system works at full power~\citep{Ranieri2009,Miele2004}. Consequently, only the direction of the thrust remains as a control parameter. A constant angle of the thrust with the instantaneous velocity vector (when accelerating and other for braking) is considered as being a compromise between a simple solution and the optimisation procedure used by~\citet{Miele2004}. This simple procedure does not assure a real optimal solution for a given mass of the mission but it should be enough to compare the performance of the propulsion systems and determine an approximate time of flight, while minimising computational time and resources.

  The duration of operations on Mars is determined by the waiting time $t_w$. Our goal is to minimise the crewed time of flight, and ultimately the total duration of the crewed part of the mission ($t_{mission} = t_{EtM} + t_{MtE} + t_w$), and for a first exploration mission an extended time for operations is not required. The waiting time is determined by the time of flight of the EtM and MtE transfers, and depends critically on the heliocentric angle between the planets upon arrival at Mars. As we show on section~\ref{sec:Results_Discussion}, for all reasonable values of the parameters of the problem, it is impossible to avoid a prolonged stay on Mars.

  % ----------------------------------------------------------------------
  %  Problem parameters
  % ----------------------------------------------------------------------
  \subsection{Problem Parameters}\label{subsec:Problem_Parameters}

  Once the baseline mission is defined (including the mass budget), and settled what is transported to Mars, we focus on the objective of trying to minimise the ToF of the crewed segment. For each propulsion system, the available thrust is increased by varying the number of engines employed.

  As argued by~\citet{Zubrin1996}\ (in the Mars Direct mission) a S/C with more than \SI{1000}{\tonne} should be avoided. However, since we aim to test new propulsion systems and establish how fast we could reach Mars and get back (and since we are not completely optimizing the continuous thrust), we stretch the limit to \SI{2500}{\tonne} for the whole mission (i.e.\ the sum of the crewed and uncrewed S/C). Without a mass limit, the mission could grow indefinitely with some ToF gains but at a prohibitive cost.

  We consider that the trajectory, and travel time, is defined by the following parameters:
  \begin{enumerate}
    \item Type of propulsion (see Table~\ref{tab:Engines_Global}), considering different number of engines to increase total thrust;
    \item Escape velocity at departure, $\vec{v}_{\infty}$ (module and direction);
    \item Thrust direction in the interplanetary phase, if applicable;
    \item Brake location during the interplanetary phase (thrust direction after brake can be different than before);
  \end{enumerate}

  For the chemical propulsion we consider a combination of two CECE engines as the unit propulsion system (i.e.\ the values of Table~\ref{tab:Engines_Global}~multiplied by two). In the case of the NERVA engine the unit propulsion system is one engine. More engines can be added to increase the available thrust and thus avoid too long working periods (that leads to higher finite burn losses).

  For the RIT-XT continuous propulsion engine we considered $5$, $10$, $15$, $25$, $35$, $45$, $60$, $75$, $100$, $150$, and $250$ engines, because of its low thrust and mass. Nevertheless, we must remember that there is an associated energy system (that increases the mass), and that more thrust does not necessarily means a better performance. In the case of the PEMT concept, we only considered combinations of $1$, $2$, $3$, and $4$ engines due to its high mass.

  For the escape velocity, we started with the one corresponding to the Hohmann transfer (HT) and searched for solutions with exponentially increasing values ($v_{\infty}=[ v^{HT}_\infty (1+0.02)^i]^2,\,i=0,1,2,3,\ldots,200$). On the return transfer, we would in principle obtain retrograde interplanetary transfer orbits, when the value of $\vec{v}_\infty$ is larger than the velocity of the planet. This case has no advantages relatively to the equivalent direct transfer, and it will not be considered.

  Regarding the angles of both the escape velocity with the velocity of the planet ($\theta_v$), and of the continuous thrust in the interplanetary phase (if applicable) with the velocity vector ($\theta_{F_T}$), we consider values from $0^\circ$ to $-90^\circ$ with a $-5^\circ$ step. The negative sign indicates that the velocity or thrust makes a retrograde angle with the velocity of the planet and the velocity of the S/C, respectively. We only considered negative angles as tests demonstrated that, all things equal, positive angles would lead to worst performance, as expected~\citep{Miele2004}.

  Braking is defined as negative thrust in the considered direction, which is equivalent to a (positive) thrust at an angle $180^\circ$ larger. Usually, reversing the thrust is only considered after the middle of the trajectory~\citep{Miele2004,Conway2010} but after testing we decided to consider that the reverse could occur at any of nine equally spaced radial distances between Earth and Mars.

  % ----------------------------------------------------------------------
  %  Results \& Discussion
  % ----------------------------------------------------------------------
  \section{Results and Discussion}\label{sec:Results_Discussion}

  % ----------------------------------------------------------------------
  %  Effects of Angle Variation
  % ----------------------------------------------------------------------
  \subsection{Effects of Angle Variation and Direction}\label{subsec:Angle_Variation}

  To better perceive the effect that $\theta_v$ and $\theta_{F_T}$ have on the system mass and ToF, we plotted the evolution of time versus mass, for an engine configuration and a single initial velocity (without brakes and for an one way trip).

\begin{figure}[!hbt]
    \centering
    \includegraphics[width=1\textwidth]{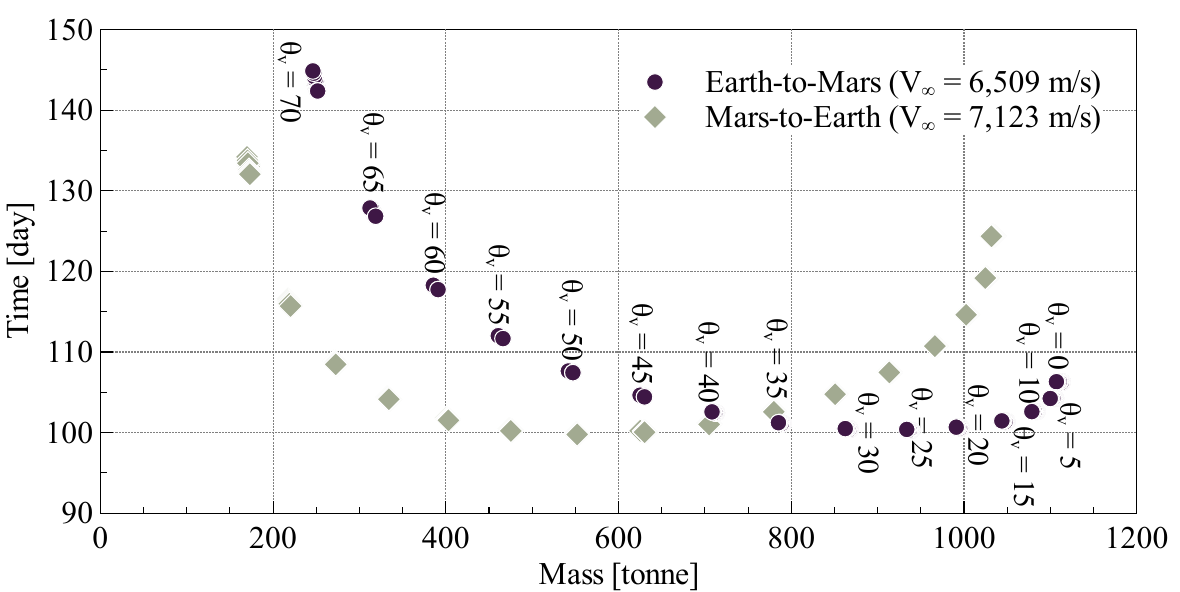}
    \caption[Variation of Velocity and Thrust Angles]{Interplanetary time of flight versus crewed total S/C mass for ten RIT-XT engines, with constant initial velocity and multiple velocity and thrust angles}
    \label{fig:Teta_Var_RIT}
  \end{figure}

  In Fig.~\ref{fig:Teta_Var_RIT} we can see how a variation on the direction of the initial velocity (maintaining the same absolute value) result in variations of mass of the order of magnitude of the hundred tonne, and tens of days. For simplicity only the $|\theta_v|$ variation are annotated in the figure, and only for an Earth to Mars transfer. The variation of $\theta_{F_T}$ does not produce as strong effect as $\theta_v$ here, although for $|\theta_v| = 70^\circ$ some dispersion of the results can be seen (which is related with $\theta_{F_T}$).

  A closer analysis of the results reveals that the cause for the mass variation is the velocity on arrival. Through the change of the initial angle we induce a variation on the interplanetary ToF and, most importantly, a different S/C velocity vector when arriving at the target planet, affecting the arrival manoeuvre (and the required propellent), and hence every manoeuvre before that.

  The fact that the thrust direction is not as important as the initial angle is explained by the relatively small continuous thrust considered in this situation. Higher thrust magnitudes induce a higher dispersion of results for the same initial velocity angle.

  An interesting fact shown by Fig.~\ref{fig:Teta_Var_RIT} is the absence of higher values of $\theta_v$. For some values of the initial velocity and thrust magnitude, the S/C cannot reach the target planet when using high values of the initial velocity angle. A classical example of this is the Hohmann transfer, for which the only allowed angle for the initial velocity is zero.

  Another aspect seen in Fig.~\ref{fig:Teta_Var_RIT} is that there are combinations of angles which yield missions with higher system mass and ToF (e.g.\ a mission with $\theta_v = 0^\circ$ has a higher mass and ToF than one with $|\theta_v| = 20^\circ$). As we aim to minimise both, these missions can be discarded. Therefore, for each escape velocity on departure we compute all $\theta_v$ and $\theta_{F_T}$ cases and select the best using a Pareto efficiency criteria (i.e.\ by ordering the cases in ascending order of time and then of mass, the points that show equal or higher times with higher masses are discarded).

  Results show that most missions selected involve braking during the interplanetary phase (less than \SI{2}{\percent} do not include a brake), being more probable for the brake to start after the middle of the transfer, in accordance to~\citet{Miele2004}.

  % ----------------------------------------------------------------------
  %  Chemical versus Nuclear Thermal propulsion
  % ----------------------------------------------------------------------
  \subsection{Chemical versus Nuclear Thermal propulsion}\label{subsec:Chemical_vs_Nuclear}

  In the different architectures we considered, an impulsive propulsion system is always required, as it is needed when performing coast interplanetary transfers, or by the continuous thrust systems for the escape and capture manoeuvres (as explained in section~\ref{subsec:Mission_Architecture_Timeline}). Therefore, the first test is to understand which of the systems (chemical or nuclear thermal) have a better mission performance for simple coast trajectories (since the continuous thrust may be seen as an addition to these solutions).

  Results show that the pure chemical CECE propulsion system performs better than the nuclear thermal NERVA propulsion, for total mission masses lower than $\sim \SI{1250}{\tonne}$. Between about \SI{1250}{\tonne} and about \SI{1600}{\tonne} both options are fairly similar. Whereas, for total mission mass higher than about \SI{1750}{\tonne} the NERVA system yields slightly better results (differences in mission mass of less than \SI{22}{\percent} of the total mission mass, for the same mission time).

  Up to the mission mass order of magnitude for which the NERVA was designed for, about \SI{1000}{\tonne}, the CECE smaller propulsion system mass outperforms the NERVA. Afterwards, the $\sim \SI{34}{\tonne}$ of the NERVA propulsion system represent less than \SI{3.4}{\percent} of the mission mass, and its higher $I_{sp}$ and thrust start to have a positive impact on the total mission mass. These results also suggest that even if the performance could be enhanced as promised by the Phoebus proposal~\citep{Sue1987}, no significant improvement would be achieved.

  Considering these results, we only used chemical propulsion for departure and capture of the continuous thrust missions. Additionally, chemical propulsion has the advantage of being simpler, and more environmentally friendly, than nuclear propulsion~\citep{Klein1970,ThePlanetarySociety2005}.

  % ----------------------------------------------------------------------
  %  Electric Engines
  % ----------------------------------------------------------------------
  \subsection{Electric Engines}\label{subsec:RD_Electric_Engines}

  For the case of electric propulsion, we tested which configuration (in terms of number of engines) yields the best results. As the different engine configurations lead to similar outcomes we decided to determine which one has the lowest Pareto curve on average (for mission masses lower than our defined limit), using a trapezoidal integration. For the RIT-XT engine the configuration with lowest value involves $25$ engines (although the difference between the minimum and maximum values is very small).

  We compare the selected RIT-XT engine with possible future technologies with increased specific impulse, $I_{sp}$, or thrust-to-weight ratio, $F_T/w$,, as explained in section~\ref{subsec:M_Electric_Engines}. For the purposes of understanding the possible gains of such enhanced electric propulsion technology, we just computed the Earth to Mars transfer time of flight of the crewed S/C. In each case we selected the best combination of electric engines (in terms of number of engines), resulting in the shortest ToF for the selected mass of the S/C. Results for a crewed S/C with \SI{1000}{\tonne} can be found in Table~\ref{tab:RITenhanced} and show that, although the considered values of specific impulse and thrust-to-weight of the enhanced technology are considerably larger, the performance change is marginal, and can even be degraded. The same result is obtained for other masses within the overall range of interest.

  \begin{table}[!htb]
    \centering
    \caption[Main Results]{Comparison of the time of flight to Mars using electric propulsion with increased specific impulse or thrust-to-weight (maintaining the remaining parameters of the engine constant, and for a wet mass at departure from Earth of the crewed S/C of \SI{1000}{\tonne})}
    \label{tab:RITenhanced}
    \begin{tabular}{lccccccc}
      \hline\noalign{\smallskip}
      Case test	& \begin{tabular}{@{}c@{}} $F_T/w$ \\  $[\si{\newton\per\kilogram}]$ \end{tabular}
        & $\frac{F_T/w}{(F_T/w)_{\text{RIT}}}$
        & $I_{sp}$ [\si{\second}]
        & $\frac{I_{sp}}{(I_{sp})_{\text{RIT}}}$
        & \begin{tabular}{@{}c@{}} N. of \\  engines\end{tabular}
        & ToF  $[\si{\day}]$
        & $\frac{\text{ToF}}{\text{(ToF)}_{\text{RIT}}}$ \\
      \noalign{\smallskip}\hline\noalign{\smallskip}
      RIT-XT (baseline)
          & \num{4.3e-3}    & 1     & \num{4600}    & 1    &  15 & 89.61 & 1	  \\
      Larger $I_{sp}$
          & \num{4.3e-3}    & 1     & \num{10000}   & 2.17 &  25 & 90.81 & 1.013 \\
      Larger $F_T/w$
          & \num{14.2e-3}   & 3.3   & \num{4600}    & 1    &  75 & 89.26 & 0.996 \\
      Both $I_{sp},F_T/w$ larger
          & \num{14.2e-3}   & 3.3   & \num{10000}   & 2.17 & 250 & 89.16 & 0.995 \\
      \noalign{\smallskip}\hline\noalign{\smallskip}
    \end{tabular}
  \end{table}

  Our interpretation of these results is that continuous thrust is not the best choice of propulsion technology for the proposed goal of minimising the ToF for this architecture. Low intensity continuous thrust requires time to be effective but to diminish the travel time is exactly the goal. As the technology improves, by improving the specific impulse or the thrust-to-weight ratio, the travel time diminishes, making the propulsion less effective and requiring more chemical propellant for the capture. This agrees with the existing literature: keeping the thrust-to-weight ratio constant and increasing the specific impulse increases the interplanetary transfer time while the interplanetary propellant mass decreases (i.e. changing the value of $I_{sp}$ has a positive effect on the mass but a negative one on the transfer time)~\citep{Miele2004}. Whereas the opposite effect for the transfer time and propellant mass is seen when the $I_{sp}$ is fixed and the thrust-to-weight is increased~\citep{Miele2004}. The surprisingly small increase in performance suggests that large gains cannot be achieved even with future technological developments of the continuous propulsion. We therefore focus our analysis on the RIT-XT engine when comparing with other types of propulsion.

  % ----------------------------------------------------------------------
  %  Nuclear Pure Electro-Magnetic Thrust
  % ----------------------------------------------------------------------
  \subsection{Nuclear Pure Electro-Magnetic Thrust}\label{subsec:RD_PEMT_Engine}

\begin{figure}[!hbt]
    \centering
    \includegraphics[width=1\textwidth]{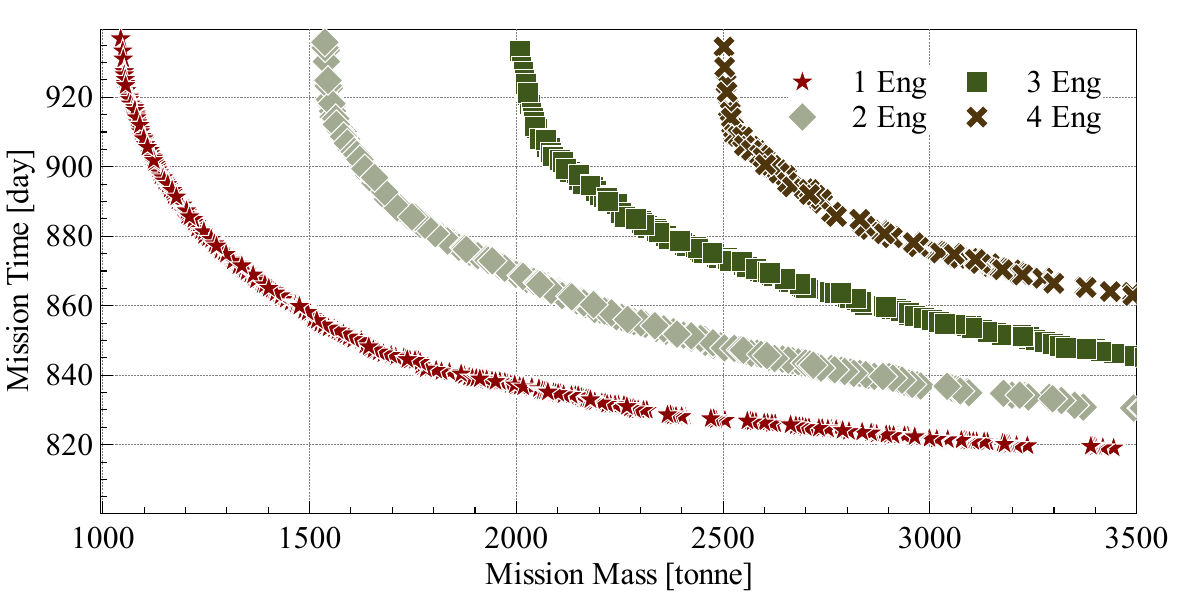}
    \caption[PEMT Mission Results]{Mission total time versus mission total mass for the PEMT propulsion for different number of engines}
    \label{fig:PEMT_comparison}
  \end{figure}

  We expected the PEMT engine to perform better than the other studied propulsion technologies, as it has a much higher $I_{sp}$ than any other engine. However, this technology has the disadvantage of a much higher permanent mass demanding much more from the auxiliary chemical propulsion for the capture. Fig.~\ref{fig:PEMT_comparison} plots the total crewed mission time versus the total mass of the mission for different configurations, and it is explicit that for the allowed mission mass range only the one engine case is worth considering. Hence, we only compare the single PEMT engine case with the other technologies. Notice that the PEMT is a continuous thrust type engine and shares some of the disadvantages of the electric engines observed in section~\ref{subsec:RD_Electric_Engines}. A larger force, with the corresponding increase in engine mass, can be disadvantageous for the defined mass bounds because it requires more chemical propellants and deposits for the capture.

  % ----------------------------------------------------------------------
  %  Impulsive versus Non-Impulsive Systems
  % ----------------------------------------------------------------------
  \subsection{Systems Comparison}\label{subsec:Impulsive_vs_Non-Impulsive}

\begin{figure}[!hbt]
    \centering
    \includegraphics[width=1\textwidth]{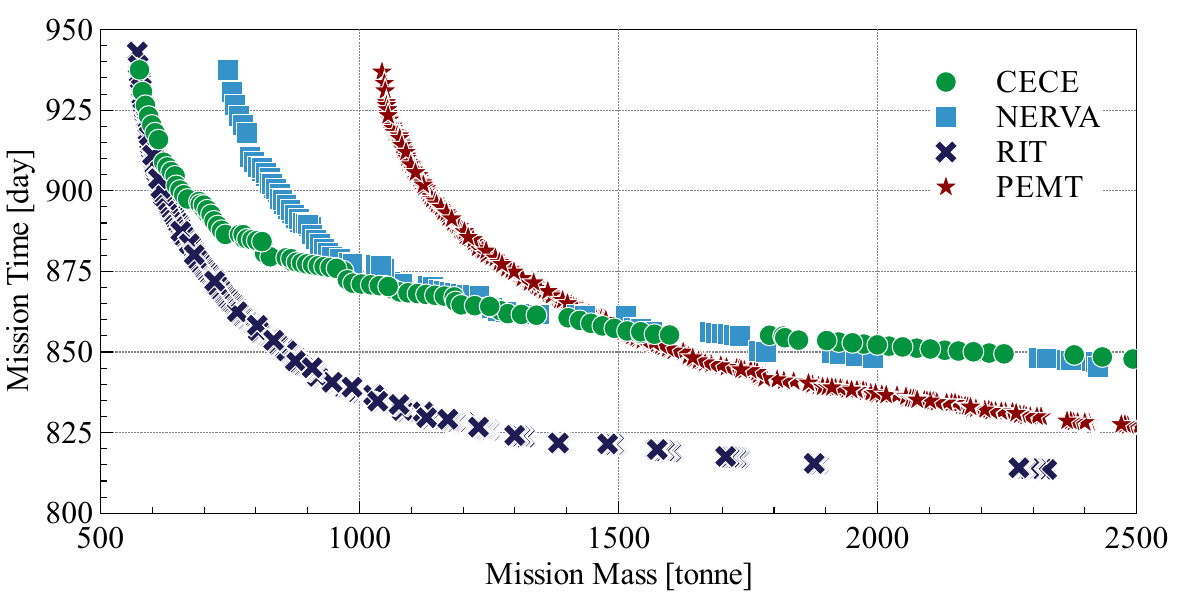}
    \caption[Global Mission Results]{Total mission time (times of flight plus waiting time) versus total mission mass (crewed plus cargo S/C) for all systems configurations}
    \label{fig:Global_Mission}
  \end{figure}

  The results of total mission time and mass for the classical chemical solution, the nuclear thermal engine, $25$ RIT-XT engines, and one PEMT, for several values of the initial escape velocity, are shown in Fig.~\ref{fig:Global_Mission}. As can be seen, the electric systems give rise to the lowest curves (i.e.\ the lowest mass for the same ToF). The exception is for mission time higher than $\sim \SI{910}{\day}$, where the simple classical chemical solution display similar outcomes (with differences less than \SI{2}{\percent} of total mission mass for the RIT-XT). Conversely, the system that shows higher mission masses is the PEMT, at least until about \SI{1500}{\tonne} of total mission mass, from which the impulsive solutions have the worst results.

  Comparing the classical chemical solution with the 25 engines electric solution, we observe that  adding to a largely optimised CECE system, a low mass and high $I_{sp}$ system, such as the electric propulsion, leads to improvements, in spite the limitations emphasised in section~\ref{subsec:RD_Electric_Engines} of the electric propulsion in increasing performance.

  When using impulsive systems, given their low $I_{sp}$, an average of about \SI{80}{\percent} of the S/C mass is propellant (here including propellants and respective tanks to transport them), for this mass range. When adding the electric system this drops to about \SI{75}{\percent}, as the electric system is introduced not only for the S/C to reach its destination faster but also for saving propellant by braking before arrival. Of this \SI{75}{\percent} propellant mass, only $\sim \SI{1}{\percent}$ corresponds to the electric system propellant.

  Combining all EtM and MtE missions (as described in section~\ref{subsec:Angle_Variation}), we determine the waiting time, total mission time and total mission mass. For the missions options presented in Fig.~\ref{fig:Global_Mission}\ the minimum waiting time achievable is $\sim \SI{520}{\day}$ (which corresponds to total mission times higher than $\sim \SI{900}{\day}$). On the other extreme (i.e. total mission masses close to the \SI{2500}{\tonne} limit and $\lesssim \SI{850}{\day}$ of mission time) the waiting time rises to $\sim \SI{570}{\day}$, for continuous thrust options, and to about \SI{660}{\day}, for the classical chemical and nuclear thermal options. The classical options show waiting times consistently higher.

  The waiting time evolution exposes its relation with travel time (and consequently the heliocentric angle). Even if we travel faster to Mars and back, the mission time is about the same, as the waiting time increases. Moreover, the waiting time is not only related to the travel time but also with $\theta_v$, i.e.\ a different $\theta_v$ requires the target planet to be in a different rendezvous location, affecting the waiting time for the return. This reveals another advantage of using continuous thrust, the change in the velocity direction that can be imposed to the S/C. In many of these mission options the thrust is not aligned with the S/C velocity but has a transverse component, which shortens the distance travelled, and affects the waiting time.

  Comparing the mission outlined here (using 25 electric propulsion engines) with the Mars Direct mission~\citep{Zubrin1996}, we show that it is possible to save about \SI{50}{\day} on the total mission time (and save $\sim \SI{60}{\day}$ on total ToF), while keeping the crewed S/C below the \SI{1000}{\tonne} limit as advocated in Mars Direct.

  By contrast, there is no PEMT options with masses lower than \SI{1000}{\tonne}, and it only shows better results for masses higher than $\sim \SI{1500}{\tonne}$ (the waiting time is almost \SI{100}{\day} lower than the chemical solution). The main reason for this is the way the engine generates energy. Since energy is generated through a nuclear reactor, which only burns \SI{4}{\percent} of the total nuclear material, the S/C is much heavier at arrival, requiring much more chemical propellent than other systems to be captured. For example, in an EtM transfer approximately \SI{16.6}{\tonne} of nuclear material is required for the interplanetary phase (ten times more than the electric option). Once the S/C reaches Mars the nuclear reactor still has $\approx\SI{16}{\tonne}$ of nuclear material (apart from the dry mass and propellant for the remaining transfers), which has to brake to be captured by Mars. Conversely, the electric option has already consumed all propellent and discarded its tanks (leaving it just with the dry mass and propellant for the remaining transfers). This has an exponential effect on previous manoeuvres and the required mass of the system. The option of getting rid of the PEMT engine after reaching Mars, and using another one that is already there to bring the crew back to Earth, is not considered due to the implied additional cost, the risk of discarding a large amount of nuclear propellant, and the increase in the mass to be transported that would make difficult for this options to be competitive.

  For masses well above the considered limit, the PEMT ends up offering the best results (values not shown here). This might suggest that for trajectories to celestial bodies farther than Mars, where the engine would be operating during a longer period (and using the same mission architecture assumed here), the PEMT engine may turn out to be the best solution.

  The total mission time is always large because even for the largest force and mission mass considered the ToF was never small enough that the heliocentric angle of the transfer would be larger than the heliocentric angle defined by the Earth during the transfer. Only then the waiting time could be small. When that does not happen any cut in the travel time will increase the waiting time.

\begin{figure}[!hbt]
    \centering
    \includegraphics[width=1\textwidth]{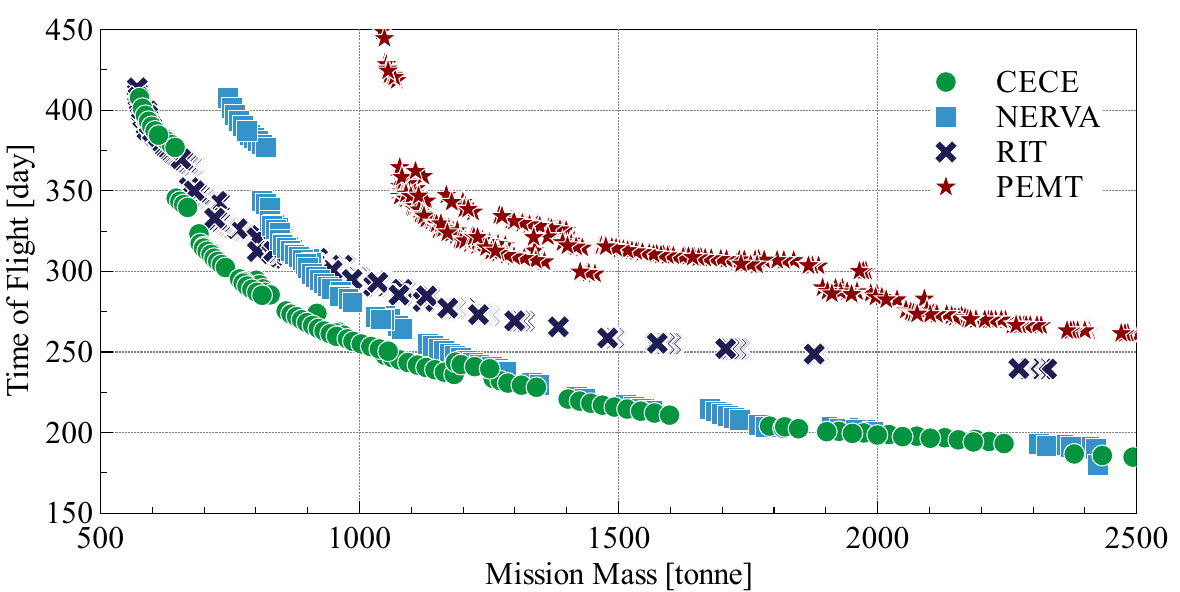}
    \caption[Global Mission Time vs Mass]{Total times of flight (Earth to Mars and return trip) for the astronauts as a function of the total mission mass}
    \label{fig:missionsTOF}
  \end{figure}

  Alternatively, instead of total mission time we could consider only the total travel time. This could be relevant if the astronauts could shelter themselves while at Mars, making pertinent just the time spent in space. Fig.~\ref{fig:missionsTOF} plots the total travel time --- from Earth to Mars and return trip of the astronauts --- against the total mission mass. In this case the classical chemical propulsion reveals to be the best, with NERVA being a very close second. When considering only the transfer times, it becomes clear that the advantage of the electric propulsion comes from saving mass and only stands because, regarding impulsive types of propulsion, the faster they are the longer they have to wait in Mars. This result suggests that chemical propulsion would be the best to try to transfer to Mars before the opposition of Mars relative to the Earth, although a much larger mass than the considered in this work would be required.

  % ----------------------------------------------------------------------
  %  Conclusions and Future Work
  % ----------------------------------------------------------------------
  \section{Conclusions}\label{sec:Conclusions}

  We have shown that it is very difficult to significantly cut the duration of a mission to Mars, including the travel times (for this mission architecture). In fact, a reduction of an order of magnitude seems impossible with the foreseeable technologies and reasonable total mission mass. Moreover, the gains in travel time end up being lost in the additional required waiting time, so a significant improvement is not to be expected.

  In the case of variable angle of thrust and use of continuous thrust in the cargo S/C an improvement in the total mass is to be expected but this would affect equally all options and a significant change is not foreseen.

  The classical chemical propulsion gives the shortest heliocentric transfer times. In most cases, for the same total ToF, it has the smallest mass.

  Electric propulsion offers some advantage, cutting the total mission mass (but not the travel time), however the advantages are limited, for the mission architecture assumed, since gains in the propulsion are lost when the travel time is cut. Our analysis suggests that an interesting combination might involve about twenty five electrical engines. However, the interaction between them should be studied, as suggested by the~\citet{NationalAeronauticsandSpaceAdministration2008}.

  The PEMT propulsion system revealed to be inadequate for a crewed mission to Mars. The main reason is that it uses only a low fraction of the total nuclear material for propulsion. If this fraction is increased, through technological improvements, the performance of the PEMT system can be greatly improved. Nevertheless, as it stands, the PEMT concept shows a better result for higher total mission mass values, suggesting that it may be more suitable for farther destinations than Mars.

  % ----------------------------------------------------------------------
  %  Acknowledgments
  % ----------------------------------------------------------------------
  \begin{acknowledgements}
    The work of Paulo J. S. Gil was supported by FCT, through IDMEC, under LAETA, project UIDB/50022/2020. The authors would like to thank John Brophy for helpful comments and suggestions on some issues of this manuscript.
    On behalf of all authors, the corresponding author states that there is no conflict of interest.
  \end{acknowledgements}

  % ---------------------------------------------
  % ---------------------------------------------
  %	REFERENCE LIST
  % ---------------------------------------------
  \bibliographystyle{spbasic}      % basic style, author-year citations
  \bibliography{Mars_Propulsion_Springer_BIB}

\end{document}